\documentclass[12pt]{article}     
\usepackage{graphics}

\newcommand{\sqrtg}{\sqrt{-\gamma}}

\newcommand{\half}{\frac{1}{2}}
\newcommand{\be}{\begin{equation}}
\newcommand{\ee}{\end{equation}}
\newcommand{\adota}{\frac{\dot{a}}{a}}
\newcommand{\veca}{\mathbf{a}}
\newcommand{\vecb}{\mathbf{b}}
\newcommand{\vecx}{{\mathbf{x}}}

\newcommand{\si}{{\sigma}}

\begin{document}

\renewcommand{\thefootnote}{\fnsymbol{footnote}}

\begin{flushright}
DAMTP-2001-43 \\ 
 (LaTeX-ed on \today )
\end{flushright}
\vskip 12pt
\begin{center}

{\large\bf 
 Dynamics and Properties of Chiral
Cosmic Strings}

\vskip 1.2cm
{\large M. Pickles\footnote{E-mail: {\tt
M.Pickles@damtp.cam.ac.uk}} and
A.C. Davis\footnote{E-mail: {\tt
A.C.Davis@damtp.cam.ac.uk}}}\\ \vskip 5pt \vskip 3pt DAMTP,
Centre for Mathematical Sciences,\\ 
Cambridge University, Wilberforce Road, Cambridge, CB3 OWA, U.K.\\
\vskip 0.3cm
\end{center}

\vskip 1.2cm

\renewcommand{\thefootnote}{\arabic{footnote}}
\setcounter{footnote}{0} \typeout{--- Main Text Start ---}

\begin{abstract}
Chiral cosmic strings naturally arise in many particle physics models,
in particular in supersymmetric theories with a D-term. These strings have
a single fermion zero mode in the core. We derive the general equation
of motion for such strings. In Minkowski space we give the self-intersections
for an arbitary varying current on the loop, showing that the self-intersection
probability is dominated by the fraction of loop with maximal charge. We 
show how to relate the charge to the fermion condensation temperature,
arguing that strings which become current carrying at formation will
automatically have a maximal charge. Any daughter loops produced are likely
to have the same charge as the parent loop. Possible models for chiral
cosmic strings are also discussed and consequences for D-term inflation
mentioned.
\end{abstract}
\vskip 0.7cm

\section{Introduction}

Recent attempts to use topological defects, and in particular cosmic strings
\cite{VS}, for structure formation have focussed on mixed scenarios including
both strings and inflation \cite{joao}. The Nambu-Goto action was used to model
the effect of the cosmic strings. This effective action results from 
the abelian Higgs model. However, in general cosmic strings formed
in supersymmetric theories have fermion zero  modes in the core
\cite{DDT}, rendering the string current-carrying. Supersymmetric
models where a U(1) gauge symmetry is broken via a Fayet-Iliopolous
D-term can lead to a period of hybrid inflation \cite{jeannerot,
mahbub}. Cosmic strings would then form at the end of inflation, and
strings formed in this way will have a single fermion zero mode in the
core, either left or right moving,  giving rise to a chiral current
\cite{DDT}. Such a string is commonly referred as a chiral string. The
effective action of such a string is not the  Nambu-Goto action, but
the chiral action \cite{CP}.

Such strings are particularly interesting because they will have
increased longevity, since the angular momentum of the charge tends
to counterbalance the tension of the loop. A loop which is completely
stabilised classically is a vorton state \cite{DS}. The existence of vortons
puts severe constraints on the underlying particle physics theory \cite{BCDT}.
Unlike superconducting strings \cite{witten}, in the case of chiral cosmic
strings the current is not electromagnetic. Indeed, the fermion zero mode 
cannot be electromagnetically coupled due to anomalies \cite{CP}.

In a previous work \cite{paper} it was shown that chiral strings
carrying maximal or near-maximal current never self-intersect, and thus
rapidly form vorton states which
remain the same size as at formation. These will in general be much larger
than the $\mathcal{O}(100)$ string widths of those produced from
ordinary superconducting cosmic strings \cite{DS}. These
vortons will scale as ordinary matter, and thus may potentially come
to dominate the matter content of the Universe. Indeed, it has previously
been shown that this leads to very stringent constraints on the underlying
particle physics theory \cite{CD}.

For general current-carrying strings, loops are characterised by
two independent quantum numbers $N$ and $Z$, both of which are conserved.
For strings with fermionic zero modes, these are associated with combinations 
of the left and right moving currents. In the chiral case there is obviously
only one independent quantum number, formed from the null current.

The equations of motion were derived in Minkowski space in \cite{paper}.
In section 2 we re-derive the equations of motion for a 
chiral cosmic string, generalising them from the form given in \cite{paper}, 
to that of a general background space. The equations are then considered in an
expanding FRW universe. So far an analytic solution has not been found, but 
our coupled equations of motion are in a form amenable to numerical
simulation.

We expect self-intersections to be the dominant route by which cosmic
string loops disappear. In \cite{paper} we numerically simulated
cosmic string loops with constant current around the loop, examining them for
self-intersections. It was found that loops with near-maximal current
generally did not self-intersect. However, it is to be expected that
the current will vary over causally disconnected regions of a loop, so
it is natural to ask how a varying current will affect these
results. We have adapted this method to examine numerically a general class of
loops with varying current in section 3.

In section 4 we discuss what value the current on the string will
take. Since the probability of the loop self-intersecting, and hence
the lifetime of the loop, is strongly dependent on the size of the
current, this is of particular importance. We find that the crucial
factor is whether the current switched on when the string was formed,
or whether it condensed at a subsequent phase transition. We then look
at several possible candidates for the fermions which are
trapped on the string. The possibilities considered are a neutrino, or a
supersymmetric combination of a gaugino and Higgsino. 

Our conclusions are discussed in section 5.

\section{Background}

We start with an effective action for chiral cosmic strings
which is invariant under space-time and worldsheet coordinate
transformations, and which constrains the string current to be
null. Such an action is given by Carter and Peter \cite{CP}.

\be
S=-\int d^2 \! \sigma \,\sqrtg
\left[m^2-\frac{1}{2}\psi^2\gamma_{ij}\phi^{,i}\phi^{,j} \right]
\label{action}
\ee
where $\gamma_{ij}$ is the metric of the string worldsheet. This is a 
generalisation of the Nambu-Goto action, containing an extra term due to
the current. 

The action is still invariant under the transformations
\be
\phi \rightarrow \tilde{\phi}(\phi) \hspace{1.5cm} \psi \rightarrow
\tilde{\psi}=\left(\frac{d\phi}{d\tilde{\phi}} \right) \psi
\ee
and we remove this by a choice of gauge. The physical current on the string
must be null for the string to be chiral. It is also conserved and
invariant under such transformations, so that it is not the Noether
current but rather $j^i=\psi\phi^{,i} \label{curr}$.

\subsection{Equations of motion in a general background}

The method used to derive the equations of motion in a general background
follows the same lines as \cite{paper}, and also \cite{BOV}.
Varying the action (\ref{action}) with respect to $\psi$ and $\phi$ we
get, respectively,
\be
\sqrtg\, \psi \, \gamma_{ij} \, \phi^{,i}  \phi^{,j}=0 
\hspace{1.5cm}
D_i(\psi^2\phi^{,i})=0
\label{null} 
\ee
\noindent
where the first equation ensures the current is null.

In 1+1 dimensions a scalar field whose gradient is everywhere
null must be harmonic, so that the second condition becomes
\be
\phi^{,i}\psi_{,i}=0
\label{psifix}
\ee
\noindent

We fix our gauge by choosing the coordinates on the string
worldsheet; the same choices as in \cite{paper} are still
valid. The timelike coordinate is $\eta=m^{-1}\phi$, and the
second coordinate q is taken to be null. The metric then takes the form
\be
\gamma_{ij}=\pmatrix{A&B\cr B&0\cr} 
\hskip 1cm \gamma^{ij}=\pmatrix{0&B^{-1}\cr B^{-1}&-AB^{-2}\cr} 
\label{gij}
\ee
and (\ref{null}) ensures that
$\frac{\partial\phi}{\partial q}=0$, so that $\psi=\psi(\eta)$, using
(\ref{psifix}).

Varying the action with respect to $x^{\rho}$, and dividing by $m^2$,
then gives :
\begin{eqnarray*}
0 & = & \left[-AB^{-2} +\psi^2 B^{-2} \right] \left(\Gamma^{\rho}_{\mu\nu}
x^{\mu}_{,q} x^{\nu}_{,q} + x^{\rho}_{,qq} \right)
+ 2B^{-1} \left(\Gamma^{\rho}_{\mu\nu}
x^{\mu}_{,q} x^{\nu}_{,\eta} + x^{\rho}_{,q\eta} \right) \\
& & - B^{-1} x^{\rho}_{,q} \partial_q(AB^{-1})
+ \psi^2 B^{-1} x^{\rho}_{,q} \partial_q(B^{-1})
\end{eqnarray*}

We fix the residual gauge freedom by choosing $A=\psi^2$. The
equations of motion then simplify to give a geodesic equation
\be
0=x^{\rho}_{,q\eta} + \Gamma^{\rho}_{\mu\nu} x^{\mu}_{,q} x^{\nu}_{,\eta}
\label{geneqns}
\ee
Since Nambu-Goto strings are just a special case of chiral cosmic
strings, it is unsurprising that this is what is obtained. For, in the
Nambu-Goto case we would get a general geodesic equation without any
gauge constraints; the absence of a chiral current is reflected in the
fact that the coordinates are pure light-cone, since $A=0$ in this case.

We look at the equations (\ref{geneqns}) in an expanding universe with
conformal time.

\begin{eqnarray}
0 & = & x^0_{,q\eta} + \adota \left[ x^0_{,\eta} x^0_{,q} +
\vecx_{,q} \cdot \vecx_{,\eta} \right] \\
0 & = & \vecx_{,q\eta} + \adota \left[ \vecx_{,\eta} x^0_{,q} + \vecx_{,q}
x^0_{,\eta} \right]
\end{eqnarray}

The gauge conditions are now

\begin{eqnarray}
A = \psi(\eta)^2 & = & a(\tau)^2 [\tau_{,\eta} \tau_{,\eta} - \vecx_{,\eta}
\cdot \vecx_{,\eta}] \\
0 & = & a(\tau)^2 [\tau_{,q} \tau_{,q} - \vecx_{,q} \cdot \vecx_{,q}]
\end{eqnarray}

In spherical polar coordinates the equations of motion become :
$$
0=\tau_{,q\eta}+ \adota \left[ \tau_{,q}\tau_{,\eta} +  r_{,q}
r_{,\eta} + r^2 \phi_{,q} \phi_{,\eta} \right]
$$
$$
0=r_{,q\eta} + \adota \left[ r_{,q} \tau_{,\eta} +  r_{,\eta}
\tau_{,q} \right] - r\phi_{,q}\phi_{,\eta}
$$
$$
0=\phi_{,q\eta} + \adota \left[ \tau_{,q}\phi_{,\eta}
+\tau_{,\eta}\phi_{,q} \right] + \frac{1}{r} \left[ r_{,q}\phi_{,\eta}
+r_{,\eta}\phi_{,q} \right] 
$$
\noindent
where $\phi$ is the azimuthal angle, and we are looking for solutions
with $\theta=\frac{\pi}{2}$. So far we have yet to find an analytic solution to
these equations. These coupled equations could be studied numerically, though
this is outside the scope of this paper.

The rest of this paper deals with chiral strings in flat spacetime.

\subsubsection{The equations in Minkowski Space}

In flat spacetime the equations of motion reduce to the wave equation 
\cite{paper}.
We use a temporal gauge
$t=\half(q+\eta)$, and define $\sigma=\half(q-\eta)$ where $\sigma$
measures equal energy intervals along the string.

The general solution is then of the form 
\be
\vecx(\eta,q)=\frac{1}{2}[\veca(t+\sigma)+\vecb(t-\sigma)].
\label{gensoln}
\ee

Examining the original gauge constraints, which relate to the
metric elements, we find that they have become
\be
\mathbf{\acute{a}}^2=1 \hspace{1.5cm} k^2 \equiv \mathbf{\acute{b}}^2
\leq 1
\label{gauge}
\ee
\noindent
where $\acute{}$ denotes differentiation with respect to the argument.

The current is $j^q=m\psi B^{-1} \hspace{0.5cm} j^{\eta}=0$ which
gives the physical current as 
\be
j^t=j^{\sigma}=\frac{m\psi}{2B}
\ee
\noindent
and the conserved charge on the string is
\be
N=\int d\sigma \sqrtg j^t=\int d\sigma m\psi(\sigma) = \frac{m}{2}
\int d\sigma \sqrt{1-k^2}
\label{charge}
\ee

Obviously there is only one independent conserved quatum number in the chiral
case since there is only a left or right moving fermion zero mode in the
string core. This is unlike the general case of a current-carrying string
where there are two conserved quantities, $N$ and $Z$, being combinations of
the left and right moving currents.

The Nambu-Goto case, which has $k=1$, thus has zero charge, as required for
consistency. However, unlike the Nambu-Goto case, for a string with
non-zero current everywhere, there are no cusps
($|\mathbf{\dot{x}}|\neq 1$). Furthermore, from the metric we see that
the physical length $l$ can be found from
\be
dl=\sqrt{\gamma_{\sigma\sigma}} d\sigma = \frac{1}{2} \left[
1+k^2- 2 \mathbf{\acute{a}} \cdot \mathbf{\acute{b}} \right]^{1/2} d\sigma.
\label{physlength}
\ee
So for $k=1$ Nambu-Goto strings $\mathbf{\acute{a}}=-
\mathbf{\acute{b}}$ and $dl=d\sigma$, while for $k=0$ 
$dl=d\sigma /2$. As we have shown previously \cite{paper}, the $k=0$ leads 
immediately to the production of stable loops or vortons.

\section{Intersection of varying current chiral strings in a flat background}

Since chiral strings have a current that tends to straighten them out,
is it still possible for them to self-intersect? In \cite{paper} we
discussed this numerically for the case of loops with
constant current in a flat background. We found that the intersection
probability was only substantially affected by near-maximal current,
and that the current needed to be larger for more wiggly loops to have
the same effect.

The case of loops with varying current
was first examined in \cite{dani}, using a specific type of loop
and current. Recall that the charge on the string is proportional to
$\sqrt{1-k^2}$. It was found
that as the number of points at which $k$ was zero, which correspond to
maximal local current, increased the probability of intersection
decreased. However, owing to the nature of the loop used, it was
impossible to determine whether it was the number of points with
maximal local current, or simply the number of regions of high current, which
determined the string self-intersection probability.

We have now carried out a general analysis for $k$ of the type
$k(\eta)=\alpha+\beta \cos(n\eta)$, with $\alpha$ and $\beta$
constants chosen so that $k(\eta)^2\leq 1$. In fact extra cosine and
sine terms can be added to $k$, with appropriate
changes to the code, but this was not examined here as the important
features are clear from the simpler case. We generated string loops 
from series of odd harmonic terms in $q$ and $\eta$, using a modified
version of the code from \cite{KS}. The highest harmonic terms
determine how 'wiggly' the string is. Note that this method generates
string loops in their centre of mass frame, so that there are no
constant terms in $\dot{\vecx}$.

The parameters chosen for
the computation were the same as in \cite{paper} and \cite{KS}, as
the results were stable to reasonable modifications. Furthermore, when
the amplitude of variation in k was small, the results were comparable
to those for constant k, as would be expected.

The results are shown below. We see that the
intersection probability is basically unchanged from the constant case
if the amplitude of $k$ is small compared to its
average. If, however, the amplitude is comparable to, or even greater
than, the average value of $k$, then in general the probability of
self-intersection decreases as the frequency increases. 

An explanation for this is that regions of high current have more
constrained motion. From (\ref{gensoln}) we have that 
\be
\dot{\vecx}(t,\sigma)=\frac{1}{2}[\acute{\veca}+\acute{\vecb}],
\hspace{1.2cm} \vecx'= \frac{1}{2}[\acute{\veca}-\acute{\vecb}].
\label{vectang}
\ee
and so $\dot{\vecx}$ and $\vecx'$ are roughly parallel when
$k^2\equiv\acute{\vecb}^2$ is small, which corresponds to high current. If the
string has points of maximal current, then the motion is exactly
tangential to the string there. For such a
string to self-intersect, either the self-intersection has to occur
between two adjacent maximal-current points, or else the string has to
have such a shape that it can intersect over larger regions. This
second type of intersection is suppressed compared to the fixed $k$
case: for, in the fixed current case there are no points of fixed
motion preventing certain types of intersection. These large-loop
intersections are also constrained, although to a slightly lesser
extent, if the current is near-maximal rather than maximal. The size
of such a 'near-maximal' current needs to be is determined by the
wiggliness of the original string, and it must be larger for more
wiggly strings. These features are clearly seen in the graphs. 

Furthermore it is to be expected that this would be more
pronounced for low-harmonic strings, the rationale being that
low-harmonic strings have fewer small-scale wiggles and so tend
to have more of the suppressed 'large' intersections. This is also
seen, if we look at the ratio
between strings with the same average current and amplitude which
intersect with low frequency $k$ to those with
high frequency. Interestingly, though perhaps not surprisingly, the
absolute decrease in the probability of intersection is about 0.05 in
all cases for a current of $k=0.1+0.1\cos(nv)$, and is similarly fixed
in other cases. This may be accounted for by assuming about 5 percent of N/N
harmonic loops for any N intersect via these 'large' loop
intersections. For $k=0.1+0.2\cos(nv)$ the figure is about 15 percent
as the minimal current is smaller, and so the loop can curve more, and
hence self-intersect on larger scales more easily. 

One other thing that stands out from the result is that the
intersection probability decreases more slowly, or even increases
between $n=2$ and $n=6$. This is probably due to the fact that in this
model the current adds higher harmonics to the string, although these
are suppressed by a factor of the magnitude of the current. Thus the
effect is substantially more pronounced for larger currents. At higher
frequencies this is negligable in comparison to the stabilising effect
of the maximal-current regions.

The only exceptions to this trend are the cases $k=0.4+0.4\cos(nv)$
and $k=0.4+0.39\cos(nv)$ for 25/25 harmonic strings. In this case the
strings are extremely wiggly on small scales, and 'larger' intersections
do not contribute substantially. 

Finally, it appears that the intersection probability is also
determined by the maximal value of $k\equiv k_{max}$, so that (at least
for small  n), the probability of intersection is close to that of the
fixed $k$ case with $k=k_{max}$.

\centerline{\resizebox{7.5cm}{!}{\includegraphics{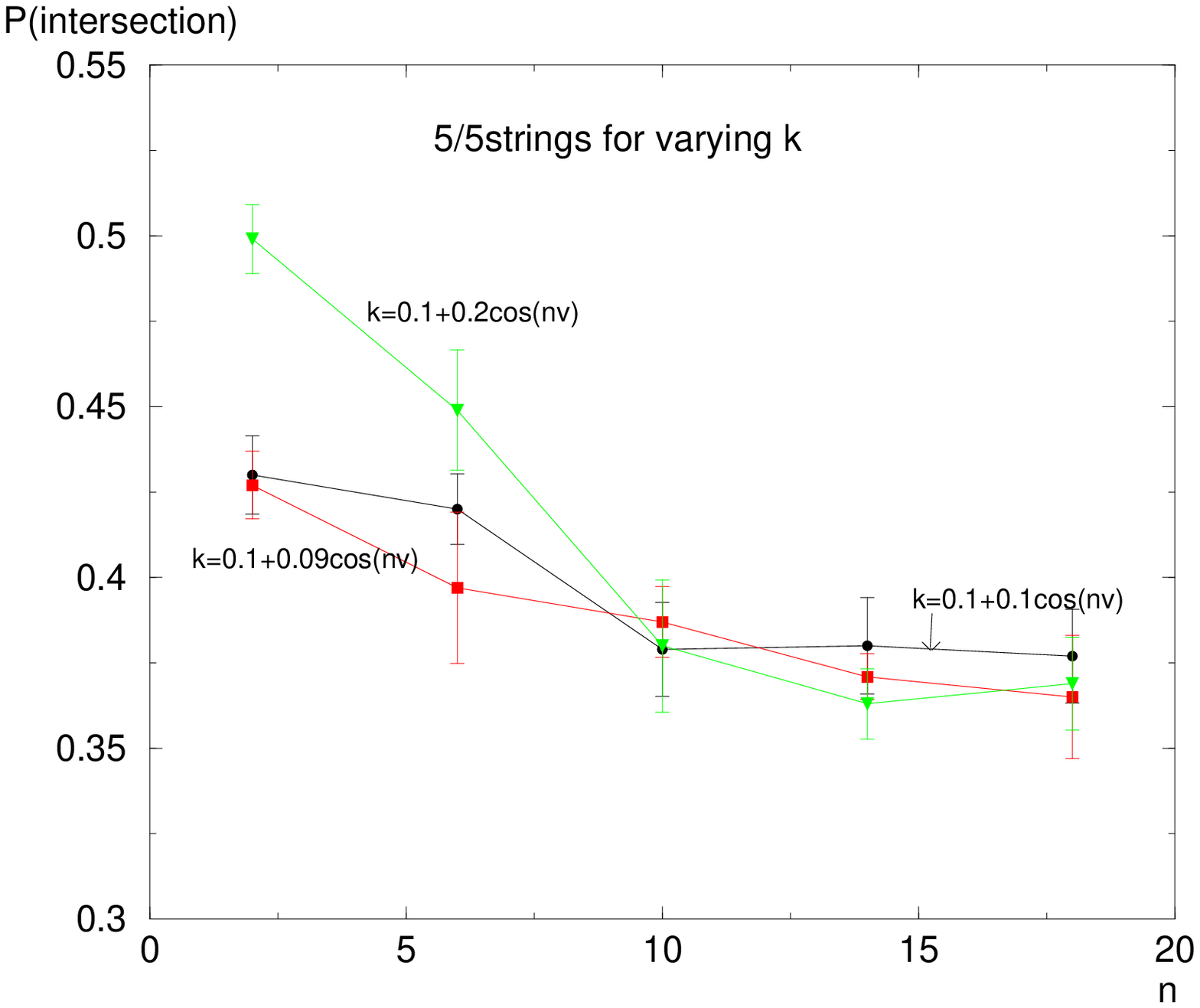}}
\resizebox{7.5cm}{!}{\includegraphics{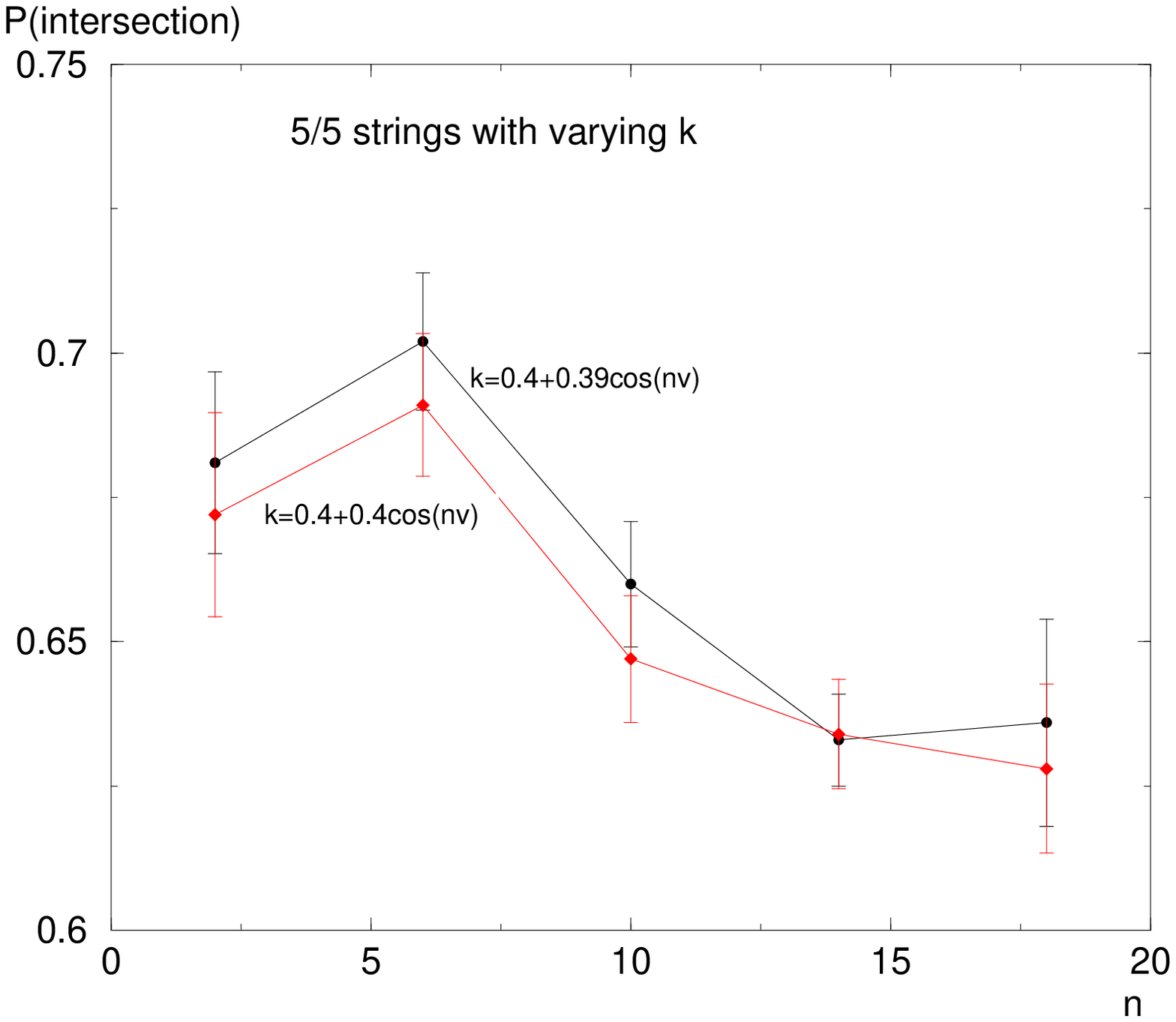}}}

\centerline{\resizebox{7.5cm}{!}{\includegraphics{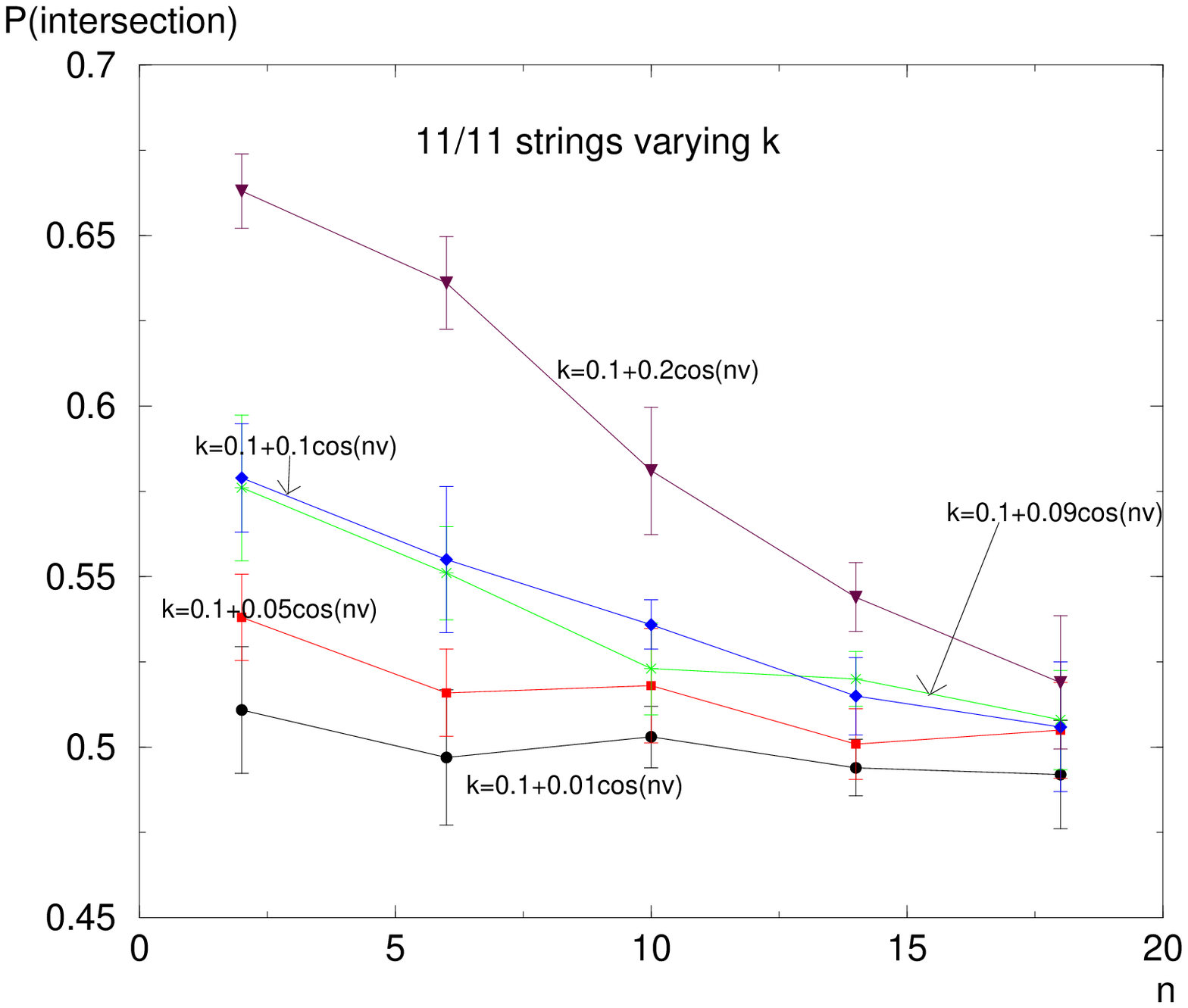}}
\resizebox{7.5cm}{!}{\includegraphics{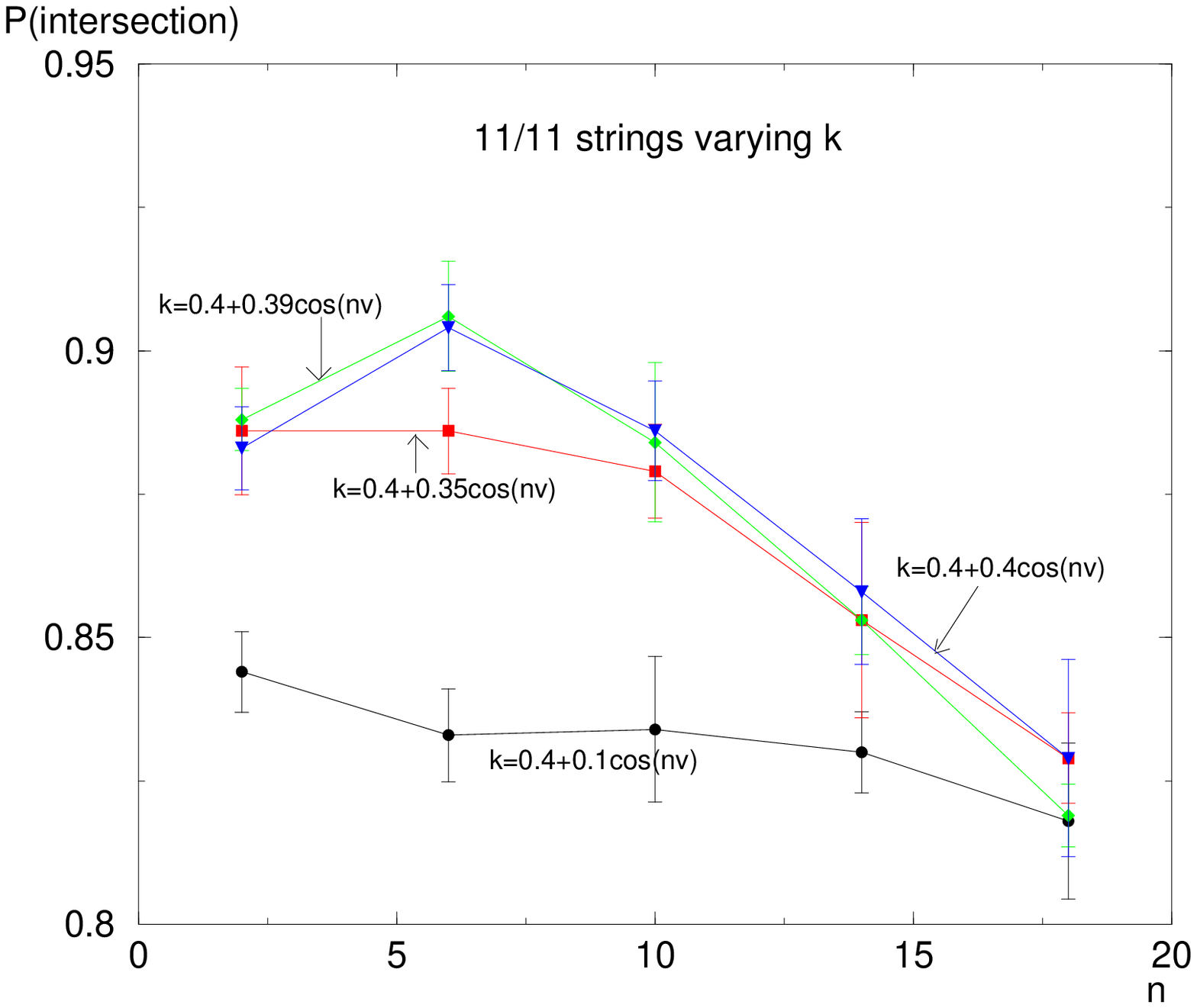}}}

\centerline{\resizebox{7.5cm}{!}{\includegraphics{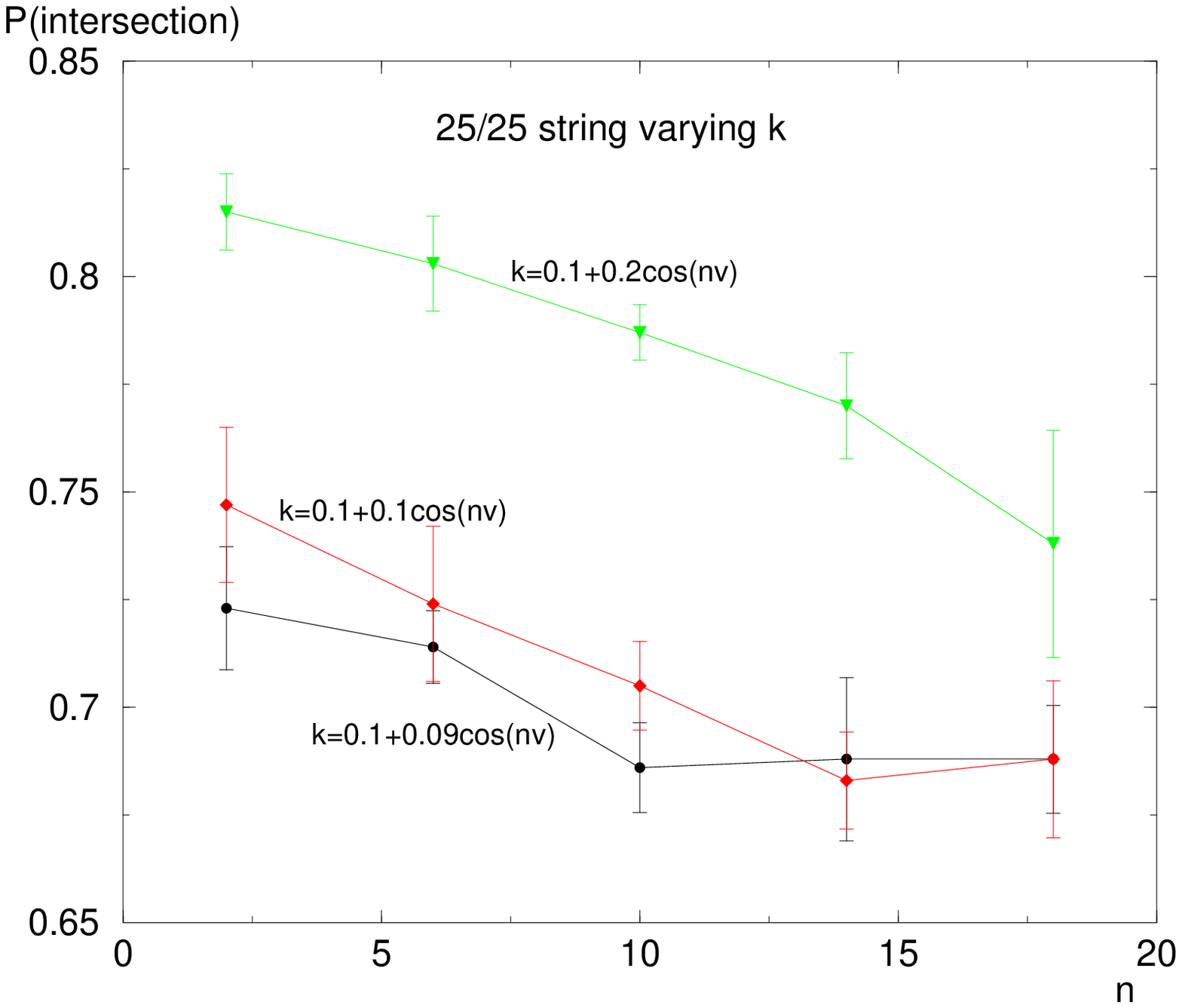}}
\resizebox{7.5cm}{!}{\includegraphics{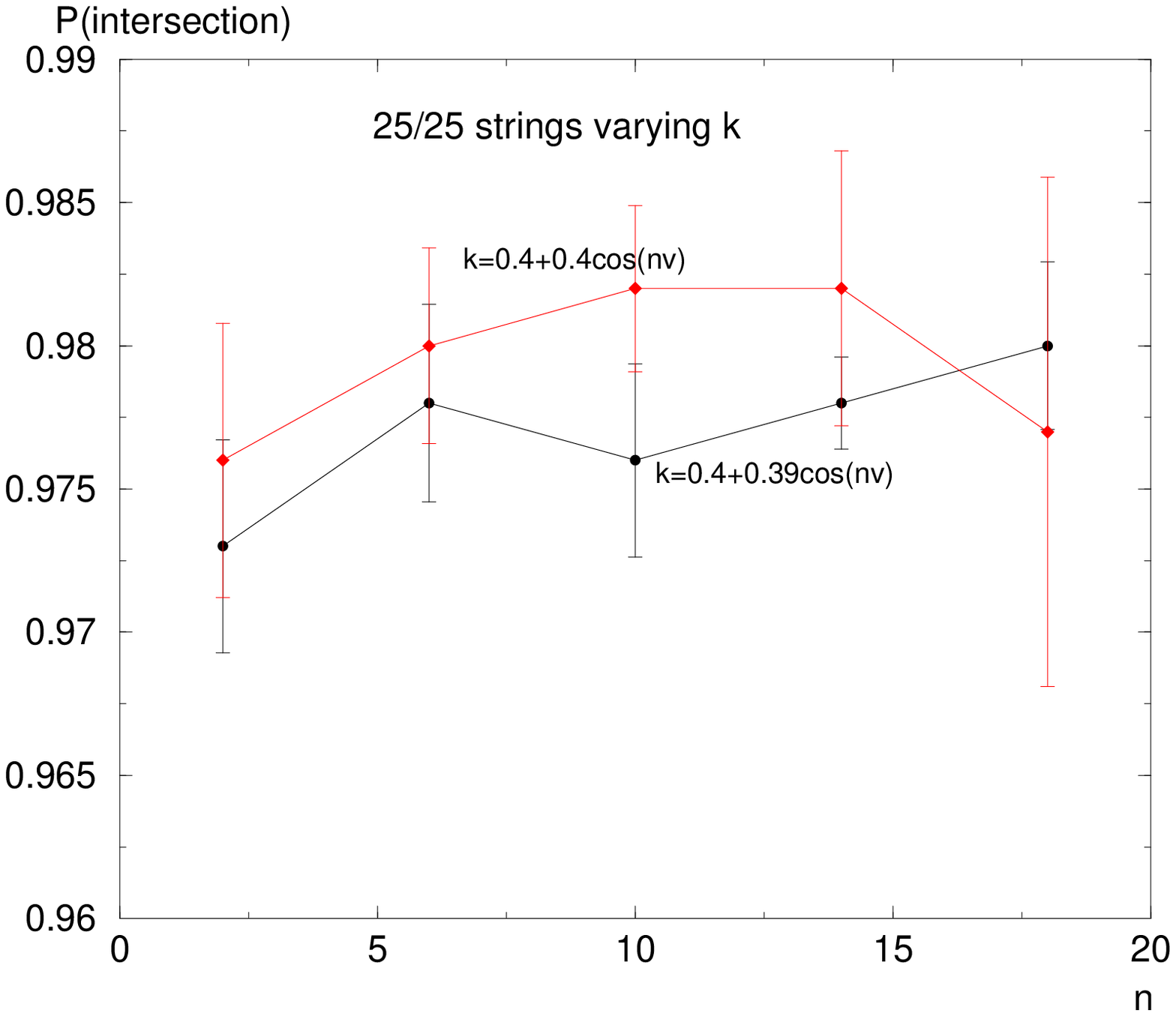}}}
\centerline{From top to bottom, varying k for 5/5, 11/11 and 25/25 strings}

\section{Properties of chiral cosmic strings}

\subsection{The current at formation}

We have not yet discussed when the current switches on, or what its
value is when it condenses. It is possible for the current to condense as
the string is formed, for instance in the case of chiral strings
resulting from D-term supersymmetric models. However, this need not be
the case, and the current may only switch on at a subsequent phase transition. 

The value of $k$ at formation, and hence the charge, is in fact
closely determined by when the current switches on.
For a chiral loop, the number of fermions $N$ on the loop will be
given by $N\approx L_{phys}/ \lambda$ \cite{CD} where $\lambda$ is the
wavelength of the fermion. If the fermion condenses at temperature
$T_{\si}$
then we expect $\lambda \sim T_{\si}^{-1}$ (in fact this should
be a very good approximation). 

Using (\ref{charge}) and (\ref{physlength}), we have the relationship
\be
N=m\int dl\sqrt{1-k^2} \left[
1+k^2-2\acute{\veca}\cdot\acute{\vecb} \right]^{-1/2}
\sim L_{phys}T_{\si}
\label{formk}
\ee
Let us consider the case of constant k for simplicity. The term
$\acute{\veca} \cdot\acute{\vecb}=k\cos\theta$ for some $\theta$
will average to roughly zero over the loop. 

If the current switches on at the string formation temperature
$T_X \approx m$, then $T_{\sigma}=T_X \approx m$, and hence we get
$$
\sqrt{1-k^2} \int dl \left[ 1+k^2 - 2k \cos\theta  \right]^{-1/2}
\sim L_{phys}
$$
The solution to this is obviously $k\approx 0$.

If the current switches on well after the string formation
temperature, so that $T_{\sigma}/m \equiv \epsilon <<1$, then we get
$$
\sqrt{1-k^2} \int dl \left[ 1+k^2 - 2k \cos\theta  \right]^{-1/2}
\sim L_{phys} \epsilon.
$$
We try an expansion $k=1-\delta$, with $\delta<<1$:
\begin{eqnarray*}
L_{phys} \epsilon & \sim & \sqrt{2\delta-\delta^2} \int dl \left[ 2-2\delta
-2(1-\delta)\cos\theta +\mathcal{O}(\delta^2 ) \right]^{-1/2} \\
& = & 
\delta^{1/2} \left( 1-\frac{\delta}{4}+ \mathcal{O}(\delta^2) \right)
(1-\delta +\mathcal{O}(\delta^2))^{-1/2}\int
\frac{dl}{\sqrt{1-\cos\theta}} \\
\end{eqnarray*}
\noindent
Note that the integral will contribute a numerical factor of order
unity. The denominator diverges like $1/\theta$ for small $\theta$,
but this is removed by the $\mathcal{O}(\delta^2)$ corrections. Hence : 

\be
\epsilon \sim \delta^{1/2} \left[ 1 + \frac{3}{4}\delta
+\mathcal{O}(\delta^2)  \right]
\ee

If the current switches on a long time after the string
forms, then the left-hand side is small, and so $k\approx 1$. Thus in
this case the current at formation is small. Of-course, the current will
build up as the loop shrinks, and vorton formation is still possible,
though the resulting vortons will be smaller than in the case where 
$k\approx 0$, and the constraints on the underlying theory less severe
\cite{CD}

\subsection{Daughter loops}

If a cosmic string intersects, it will normally be expected to produce
daughter loops (eg see \cite{VS}) as the strings may intercommute while
crossing. This need not be the case, for if there is a topological
obstruction to them reconnecting the strings will entangle,
with another string forming to join them together. In that
case the strings behave very differently, and they will not be considered
further here. Note that intercommutation is necessary to obtain the usual
string scaling solution.

In general each daughter loop thus formed  will possess a kink at the
original point of intersection, owing to the 
discontinuities in the derivatives of $\veca$ and $\vecb$. These kinks
will each split  into two kinks moving in opposite directions along
the string, the right-hand one moving at the speed of light, while the
left-mover travels at subluminal velocity for $k<1$.
Owing to the rapid acceleration of charge at a kink, it is anticipated
that the kinks will quickly evaporate due to radiation, so that in a sense the
current smooths out the functions $\veca$ and $\vecb$. 

In the chiral case, the obvious question is: how is the current on the
daughter loop related to that of its parent? It seems likely that the
current on the daughter loops would be similar to that of their
parent. For, the current is determined by the value of $k$ on the
loop, which is itself determined by the shape of the loop. Just after
the formation of the daughter loops, each daughter will have the same
shape as the region of the parent loop from which it came, as this is
required by causality. The only difference is that the period of the
daughter loop is different from that of its parent. Thus, apart from
the kink region, the daughter loops will each evolve according to the
equations of motion (\ref{geneqns}) subject to the gauge
conditions (\ref{gauge}), with the initial conditions set when the
daughter loop forms. The left and right-moving modes, $\veca$ and $\vecb$,
evolve independently and the amplitude of $\vecb'$, $k$, is
unchanged. The only problem is at the kinks, but these should be smoothed
away by radiation. Consequentially the value of $k$ around a kink,
which may be discontinuous before, should be averaged out. Of course,
as the daughter loop shrinks, the current on the loop will increase since
the charge, $N$, is conserved.

\subsection{The fermion on the string}

So far we have just specified that there is a single fermion zero mode
trapped in the string core of the chiral string. What exactly could
this fermion be? 

In the N=1 supersymmetric model with a U(1) gauge
symmetry the field giving rise to the string is that of the complex
scalar part of the primary charged chiral superfield. By considering
how the fermionic sector of the theory transforms under supersymmetry
it can be shown \cite{DDP} that the zero mode is a combination of a
gaugino and Higgsino. In this case, the current forms at the same time
as the string, and by the above argument the current will be large.
Typically this is the case in D-term inflation \cite{mahbub}, so the
resulting cosmic strings are in fact chiral strings. Indeed, the strings
form at the end of inflation \cite{jeannerot}, so this class of models are 
very likely to have a vorton problem, unless the scale of symmetry breaking is
sufficiently small \cite{CD}. It might be thought that the zero modes may
not survive supersymmetry breaking. However, it was shown in \cite{DDT2}
that the chiral zero mode does survive, since there are no 
other zero modes for it to mix with to become a bound state. 

Another possibility is a neutrino zero mode. Neutrinos have very small
masses, which are usually explained in terms of the see-saw
mechanism. Right-handed neutrino singlet states are introduced,
and they are each given  a large Majorana mass $M_R$, while the
left-handed neutrinos have zero Majorana mass. This mass would be
expected to arise from a Grand Unified Theory undergoing a phase
transition, though it can also be put in by hand into the Standard
Model. There is also a Dirac mass term $M_D$, which can arise in the
usual way from coupling to the Standard Model Higgs. It can then be
shown that the left-handed neutrinos gain a mass of
order $m_{\nu}=M_D\frac{1}{M_R}M_D^T$ in matrix form for the three
fermion families. The neutrino current could condense on the string at a phase
transition after the string formation in this case.

In $SO(10)$ GUTs there are both cosmic strings and right-handed neutrinos,
which are the only fermions acquiring a mass from the breaking of the 
$SO(10)$ symmetry. Above the electroweak scale it may be
shown, using the index theorem of \cite{DDP2}, that each of the
right-handed neutrinos will contribute zero modes to the
string, and the string is a chiral string.
However, at the electroweak phase transition the
$\nu_R$ mix with the usual $\nu_L$, giving them small masses as above.
The zero modes then cease to be zero modes, becoming instead low-lying bound 
states. Again this can be seen using the index theorem \cite{DDP2}.
In this case the string would no longer be a chiral string.
However, it is possible for one of the $\nu_L$ to be strictly massless after 
the electroweak phase transition, and furthermore to be a flavour 
eigenstate so that it does not mix with the other neutrinos. In this case
the corresponding
$\nu_R$ would remain a zero mode and the string a chiral string.
The underlying GUT theory is subject to the constraints in \cite{CD}.

\section{Discussion}

In this paper we have extended our study of chiral cosmic strings and
their properties. Although string solutions in an expanding Universe
have not yet been found, this is to be expected as the same problem
occurs in Nambu-Goto strings. Nevertheless the equations of motion
have been presented in a way that will hopefully be tractable
numerically, if not analytically.

We have presented a heuristic argument for the size of the
current. In fact, this provided a strong motivation for saying
that the current is near-constant, even across causally disconnected
regions of the string, for at each point on the string the current is
determined by when the current switched on in relation to the string
being formed at that point. Furthermore, the frequency of oscillation
of the current will depend on the size of the causally connected
regions, where here the timescale is that of the condensation of the
current on the string.

It was also found that having a current that can vary around the string
loop does not change the fundamental conclusion that for near-maximal
currents the strings are not expected to self-intersect, while for
lower currents, such self-intersections are extremely likely. Thus the
idea of large vortons remains, and indeed the over-production of such
things, which would come to dominate the energy density of the
Universe, provides a constraint on all models that predict such chiral strings.
This suggests that models based on D-term inflation, with chiral strings
formed at the end of inflation, are strongly constrained by the overproduction
of vortons and are unlikely to be viable models for structure formation.

Finally we looked at how these chiral strings might arise. The most
convincing mechanism is that of a string formed by D-term SUSY
breaking, as in this case the current is expected to be large,
resulting in vorton production. Vortons may also have a role to
play in ruling out neutrino models, although this is more tentative at
the moment.

\section{Acknowledgements}
This work is supported in part by PPARC and an ESF network. We are grateful
to Tom Kibble and Rachel Jeannerot for discussions.

\end{document}